\documentclass[12pt]{article}
\usepackage{amsmath,amssymb,amsfonts}
\usepackage[dvips]{graphicx}
\usepackage{epsfig}

\makeatletter \@addtoreset{equation}{section} \makeatother

\renewcommand{\baselinestretch}{1.2}
\begin{document}

\renewcommand{\thefootnote}{\alph{footnote}}

\begin{titlepage}

\begin{center}
\hfill {\tt Imperial/TP/09/SK/02}\\
\hfill{\tt TIFR/TH/09-19}\\
\hfill {\tt arXiv:0906.4751}

\vspace{1.5cm}

{\large\bf 
Aspects of Monopole Operators in $\mathcal{N}\!=\!6$ Chern-Simons Theory}

\vspace{2cm}

\renewcommand{\thefootnote}{\fnsymbol{footnote}}

{Seok Kim$^{1}$ and Kallingalthodi Madhu$^2$}

\vspace{1cm}

\textit{\hspace*{-0.1cm} $^1$Theoretical Physics Group, Blackett
Laboratory,
Imperial College, London SW7 2AZ, U.K.}\\

\vspace{0.2cm}


\textit{\hspace*{-0.1cm} $^1$Institute for Mathematical Sciences,
Imperial College, London SW7 2PG, U.K.}\\

\vspace{0.2cm}

\textit{\hspace*{-0.1cm} $^1$Department of Physics and Astronomy \& Center for
Theoretical Physics,\\
Seoul National University, Seoul 151-747, Korea.\footnote{Address from 1 September 2009.}}\\

\vspace{0.2cm}

\textit{\hspace*{-0.1cm} $^2$Department of Theoretical Physics,
Tata Institute of Fundamental Research,\\
Homi Bhabha Road, Mumbai 400005, India.}\\

\vspace{0.7cm}

E-mails: {\tt skim@phya.snu.ac.kr, kmadhu@theory.tifr.res.in}

\end{center}

\vspace{1cm}

\begin{abstract}

We study local operators of $U(N)\times U(N)$ $\mathcal{N}\!=\!6$
Chern-Simons-matter theory including a class of magnetic monopole
operators. To take into account the interaction of monopoles and
basic fields for large Chern-Simons level $k$,
we consider the appropriate perturbation theory in $\frac{1}{k}$
which reliably describes small excitations around protected chiral operators.
We also compute the superconformal index with some simple monopole operators
and show that it agrees with the recent result obtained from localization.
For this agreement, it is crucial that excitations of gauge fields and some
matter scalars mix, which is described classically by odd dimensional
self-duality like equations.

\end{abstract}

\end{titlepage}

\renewcommand{\thefootnote}{\arabic{footnote}}
\setcounter{footnote}{0}

\renewcommand{\baselinestretch}{1}

\tableofcontents

\renewcommand{\baselinestretch}{1.2}

\section{Introduction}

Recent studies of AdS$_4$/CFT$_3$ states that a class of Chern-Simons-matter theories provide holographic descriptions of M-theory. See, among others, \cite{Schwarz:2004yj,Bagger:2006sk,Gustavsson:2007vu,Aharony:2008ug}.
An important ingredient for understanding M-theory in this setting is the magnetic monopole operator
\cite{Borokhov:2002ib}, creating the gauge theory duals of
Kaluza-Klein modes along the `eleventh direction' beyond type IIA backgrounds.
For recent studies of monopole operators, see
\cite{Berenstein:2008dc,Klebanov:2008vq,Imamura:2009ur,
Kim:2009wb,SheikhJabbari:2009kr,Benna:2009xd,Gustavsson:2009pm,Kwon:2009ar}.
The $\mathcal{N}\!=\!6$ Chern-Simons-matter theory
\cite{Aharony:2008ug} is an ideal model to study this subject.

Generically, operators (or states via radial quantization) involving monopole operators can be
understood only after taking large interactions into account even if
the coupling constant $\frac{1}{k}$ is small, where $k$ is the
Chern-Simons level. A simple argument goes as follows. For
simplicity, consider a subsector of $U(N)\times U(N)$
$\mathcal{N}\!=\!6$ theory consisting of the gauge fields
$A_\mu,\tilde{A}_\mu$ and complex scalars $\phi_I$. The argument
generalizes to other theories. The action takes the form
\begin{equation}
  \int{\rm tr}\left[\frac{k}{4\pi}\left(AdA-\frac{2i}{3}A^3\right)-\frac{k}{4\pi}
  \left(\tilde{A}d\tilde{A}-\frac{2i}{3}\tilde{A}^3\right)
  -D_\mu\phi_I D^\mu\bar\phi^I-\frac{1}{k^2}V(\phi)\right]\ ,
\end{equation}
where $V(\phi)$ is a potential which is of sixth order in $\phi_I$.
The equations of motion for $A_\mu,\tilde{A}_\mu$ are
\begin{equation}\label{schematic-gauss}
  \frac{k}{2\pi}\star F_\mu=i\left(D_\mu\bar\phi^I\phi_I
  -\bar\phi^ID_\mu\phi_I\right)\ ,\ \
  \frac{k}{2\pi}\star\tilde{F}_\mu=i\left(\phi_ID_\mu\bar\phi^I
  -D_\mu\phi_I\bar\phi^I\right)\ .
\end{equation}
With magnetic monopoles, the left hand sides on a spatial 2-sphere integrate to
$\mathcal{O}(k)$ numbers due to the flux quantization, which
requires the scalars to be $\mathcal{O}(k^{\frac{1}{2}})$. With
scalars at this order, one generically finds that both the
kinetic term and the potential $\frac{1}{k^2}V(\phi)$ contribute to the scalar
equation of motion in same orders, implying that
conventional perturbative approach cannot be valid.\footnote{An
important exception is the case in which $V(\phi)$ vanishes, which
corresponds to chiral operators. We shall consider these exceptional
cases as well in this paper.}

In this paper, we investigate the correct perturbation
theory at large $k$ appropriate for studying the spectrum of these operators. We start
by studying the classical field theory. We first find the lowest
energy configurations solving all equations of motions in the
interacting theory including (\ref{schematic-gauss}), for a class of
magnetic fluxes turned on. See section 2 for the details on the
monopole operators we consider. By studying the fluctuation of
fields in this exact background, we find the desired perturbation
theory in $\frac{1}{k}$. In this paper, we mainly study the `free
theory' limit ignoring subleading interactions, leaving more
elaborate study as a future work. Quantizing the modes in the free
theory, one can immediately calculate the partition function for these
operators.

The spectrum that we obtain in this free theory is subject to change
as one considers interactions suppressed by
$\frac{1}{k}$. However, spectrum of certain states
preserving supersymmetry can be stable against interaction at
least in the 't Hooft limit, which takes $N,k\rightarrow\infty$ with
$\lambda=\frac{N}{k}$ finite \cite{Bhattacharya:2008bja}. Such
states are counted by the superconformal index
\cite{Kinney:2005ej}.
Recently the superconformal index for the $\mathcal{N}\!=\!6$
Chern-Simons-matter theory has been computed and studied
\cite{Kim:2009wb} in the sectors containing monopole operators,
generalizing the earlier work \cite{Bhattacharya:2008bja}. In
\cite{Kim:2009wb}, the index is computed by applying localization
technique to the path integral for the index. This calculation
involves deforming the theory in a way that
the index is not changed. Although this is a standard method,
it should be illuminating if one can understand the same quantity
directly from the $\mathcal{N}\!=\!6$ Chern-Simons-matter theory
itself, at least for large $k$. We use our free theory to reproduce
this index for a class of monopoles.

Monopole operators studied in \cite{Kim:2009wb} can be
classified into two classes, according to their representations of
$U(N)\times U(N)$ gauge group. Firstly, a monopole operator can be
in a conjugate representation of the two $U(N)$ gauge groups. Such
monopole operators can combine with basic field operators to form
gauge invariant chiral operators, in the sense that theirs dimensions are
given by the R-charge. These chiral
operators are relatively well understood \cite{Berenstein:2008dc,
Klebanov:2008vq,SheikhJabbari:2009kr,Benna:2009xd}.
It is also in this sector that we can find exact classical solutions
with lowest energy. Secondly, it was shown
in \cite{Kim:2009wb} that monopole operators in non-conjugate
representations of two $U(N)$ \textit{should} exist.
This claim was solidly supported by a very detailed matching
between the large $N$ indices from gauge theory and supergravity.
In this case, the lowest energy states come with nonzero
spatial angular momenta, which is harder to study than the chiral
operators in the first class. We do not know yet how to analyze this sector
using the approach of this paper, and leave this problem as a future work.

The rest of this paper is organized as follows. In section 2 we provide the
classical analysis of the field theory on $S^2\times\mathbb{R}$,
in the presence of magnetic fluxes. By studying all small fluctuations
of charged modes and quantizing them, we calculate the superconformal index,
which agrees with the result of \cite{Kim:2009wb}. We also briefly discuss
open problems with monopole operators which are not considered in detail
in this paper. In section 3 we conclude with a few comments.

\hspace*{-0.6cm}{\bf Note:} While we were finalizing this draft,
\cite{Berenstein:2009sa} appeared with some overlap with our work.

\section{Spectrum with monopole operators}

The $\mathcal{N}\!=\!6$ Chern-Simons-matter theory is often
conveniently described by decomposing the fields in
$\mathcal{N}\!=\!2$ supermultiplets. In particular, this is useful for
us since we shall consider the superconformal index of this theory
in section 3, which uses $\mathcal{N}\!=\!2$
supersymmetry only. In the notation of \cite{Kim:2009wb} (which closely
follows \cite{Benna:2008zy}), the matter fields decompose to four
chiral multiplets $(A_a,\psi_{a\alpha})$,
$(B_{\dot{a}},\chi_{\dot{a}\alpha})$ (with $a,\dot{a}=1,2$) in bifundamental and anti-bifundamental representation of $U(N)\times U(N)$, respectively. Table 1 summarizes the global charges of fields. $h_{1,2,3}$ are three Cartans of $SO(6)$ R-symmetry, and $h_4$ is the `baryon-like' $U(1)_b$ charge. The
action on $\mathbb{R}^{2+1}$ is given as follows:
\begin{equation}
  \mathcal{L}=\mathcal{L}_{CS}+\mathcal{L}_{m}\ ,
\end{equation}
where the Chern-Simons term is given by
\begin{equation}\label{cs-action}
  \mathcal{L}_{CS}=\frac{k}{4\pi}
  {\rm tr}\left(A\wedge dA-\frac{2i}{3}A^3
  +i\bar\lambda\lambda-2D\sigma\right)-\frac{k}{4\pi}
  {\rm tr}\left(\tilde{A}\wedge d\tilde{A}
  -\frac{2i}{3}\tilde{A}^3
  +i\bar{\tilde\lambda}\tilde\lambda-2\tilde{D}\tilde\sigma\right)\ ,
\end{equation}
and (overbar for scalars denotes Hermitian conjugation)
\begin{eqnarray}\label{matter-action}
  \hspace*{-0.5cm}\mathcal{L}_{m}&=&
  {\rm tr}\left[\frac{}{}\right.\!\!
  -D_\mu\bar{A}^aD^\mu A_a-D_\mu\bar{B}^{\dot{a}}
  D_\mu B_{\dot{a}}-i\bar\psi^a\gamma^\mu D_\mu\psi_a-i\bar\chi^{\dot{a}}
  \gamma^\mu D_\mu\chi_{\dot{a}}\nonumber\\
  &&\hspace{0.5cm}
  -\left(\sigma A_a-A_a\tilde\sigma\right)
  \left(\bar{A}^a\sigma-\tilde\sigma\bar{A}^a\right)
  -\left(\tilde\sigma B_{\dot{a}}-B_{\dot{a}}\sigma\right)
  \left(\bar{B}^{\dot{a}}\tilde\sigma-\sigma\bar{B}^{\dot{a}}
  \right)\nonumber\\
  &&\hspace{0.5cm}
  +\bar{A}^aDA_a-A_a\tilde{D}\bar{A}^a-B_{\dot{a}}D\bar{B}^{\dot{a}}
  +\bar{B}^{\dot{a}}\tilde{D}B_{\dot{a}}\nonumber\\
  &&\hspace{0.5cm}-i\bar\psi^a\sigma\psi_a+i\psi_a\tilde\sigma\bar\psi^a
  +i\bar{A}^a\lambda\psi_a+i\bar\psi^a\bar\lambda A_a
  -i\psi_a\tilde\lambda\bar{A}^a-iA_a\bar{\tilde\lambda}\bar\psi^a\nonumber\\
  &&\hspace{0.5cm}+i\chi_{\dot{a}}\sigma\bar\chi^{\dot{a}}
  -i\bar\chi^{\dot{a}}\tilde\sigma\chi_{\dot{a}}
  -i\chi_{\dot{a}}\lambda\bar{B}^{\dot{a}}
  -iB_{\dot{a}}\bar\lambda\bar\chi^{\dot{a}}
  +i\bar{B}^{\dot{a}}\tilde\lambda\chi_{\dot{a}}
  +i\bar\chi^{\dot{a}}\bar{\tilde\lambda}B_{\dot{a}}\!\left.\frac{}{}\right]
  +\mathcal{L}_{\rm sup}\ .
\end{eqnarray}
$\mathcal{L}_{\rm sup}$ contains scalar potential and Yukawa
interaction obtained from a superpotential
\begin{equation}
  W=-\frac{2\pi}{k}\epsilon^{ab}
  \epsilon^{\dot{a}\dot{b}}{\rm tr}(A_aB_{\dot{a}}A_bB_{\dot{b}})\ .
\end{equation}
$\sigma,\lambda,D$ and $\tilde\sigma,\tilde\lambda,\tilde{D}$ are auxiliary.
The Lagrange multipliers $D,\tilde{D}$ impose
\begin{equation}
  \sigma=\frac{2\pi}{k}\left(A_a\bar{A}^a-\bar{B}^{\dot{a}}B_{\dot{a}}\right)
  \ ,\ \ \tilde\sigma=\frac{2\pi}{k}\left(\bar{A}^aA_a-
  B_{\dot{a}}\bar{B}^{\dot{a}}\right)\ .
\end{equation}
$\lambda_\alpha,\tilde\lambda_\alpha$ are also given in terms of the
matter fields as
\begin{equation}
  \lambda=\frac{4\pi}{k}\left(\bar\chi^{\dot{a}}B_{\dot{a}}-A_a\bar\psi^a
  \right)\ ,\ \ \tilde\lambda=\frac{4\pi}{k}\left(B_{\dot{a}}\bar\chi^{\dot{a}}
  -\bar\psi^aA_a\right)\ .
\end{equation}
This theory can be put on $S^2\times\mathbb{R}$ via radial quantization.
See, say, \cite{Kim:2009wb} for the details. One of the important
changes is that scalars acquire masses proportional to the
curvature of $S^2$. Without losing generality, we set the radius of
$S^2$ to $1$. Then one finds $m^2=\frac{1}{4}$ for all scalars.

In the radially quantized theory, one can consider configurations in
which nonzero magnetic flux on spatial $S^2$ is turned on. From the
representations of matter fields under $U(N)\times U(N)$, one finds
that ${\rm tr}F={\rm tr}\tilde{F}$ should be satisfied. The
Kaluza-Klein momentum in the dual M-theory along the fiber circle of
$S^7/\mathbb{Z}_k$ is given by
\begin{equation}\label{kk-momentum}
  P=\frac{k}{4\pi}\int_{S^2}{\rm tr}F=
  \frac{k}{4\pi}\int_{S^2}{\rm tr}\tilde{F}
\end{equation}
in the gauge theory \cite{Aharony:2008ug}. This, via Gauss' law constraint,
turns out to be proportional to $h_4$ in Table 1.
Monopole operators create these fluxes on $S^2$. In particular, we can embed
$U(1)^N\times U(1)^N$ Dirac monopoles to $U(N)\times U(N)$ such that
\begin{equation}
  \frac{1}{4\pi}\int_{S^2}F=\frac{1}{2}{\rm diag}(n_1,n_2,\cdots,n_N)\ ,\
  \ \frac{1}{4\pi}\int_{S^2}\tilde{F}=\frac{1}{2}{\rm diag}
  (\tilde{n}_1,\tilde{n}_2,\cdots,\tilde{n}_N)\ .
\end{equation}
Here $H\equiv(n_1,n_2,\cdots,n_N)$ and $\tilde{H}\equiv(\tilde{n}_1,
\tilde{n}_2,\cdots,\tilde{n}_N)$ are non-increasing
integers.

\begin{table}[t]\label{charges}
$$
\begin{array}{c|ccc|cc|c}
  \hline{\rm fields}&h_1&h_2&h_3&j_3&\epsilon& h_4\\
  \hline(A_1,A_2)&(\frac{1}{2},-\frac{1}{2})&(\frac{1}{2},-\frac{1}{2})&
  (-\frac{1}{2},-\frac{1}{2})&0&\frac{1}{2}&\frac{1}{2}\\
  (B_{\dot{1}},B_{\dot{2}})&(\frac{1}{2},-\frac{1}{2})&(-\frac{1}{2},
  \frac{1}{2})&(-\frac{1}{2},-\frac{1}{2})&0&\frac{1}{2}&-\frac{1}{2}\\
  (\psi_{1\pm},\psi_{2\pm})&(\frac{1}{2},-\frac{1}{2})&(\frac{1}{2},
  -\frac{1}{2})&(\frac{1}{2},\frac{1}{2})&\pm\frac{1}{2}&1&\frac{1}{2}\\
  (\chi_{\dot{1}\pm},\chi_{\dot{2}\pm})&(\frac{1}{2},-\frac{1}{2})&
  (-\frac{1}{2},\frac{1}{2})&(\frac{1}{2},\frac{1}{2})&\pm\frac{1}{2}&1
  &-\frac{1}{2}\\
  \hline A_\mu,\tilde{A}_\mu&0&0&0&(1,0,-1)&1&0\\
  \hline
\end{array}
$$
\caption{charges of fields}
\end{table}

\subsection{Classical solutions}

In this paper, we mainly consider monopole operators
with $n_1\!=\!\tilde{n}_1\!=\!n\!>\!0$ and other $n_i,\tilde{n}_i\!=\!0$.
Sectors with negative $n$ can be obtained by parity.
In subsection 2.3, we discuss the technical complications that we
encounter for other cases, including the cases with $H\neq\tilde{H}$.

The matter fields carry positive scale dimensions and nonzero
$U(1)_b$ charge $h_4$, where the latter charge has to be
balanced with the total magnetic flux through the Gauss' law. The
total $U(1)_b$ charge is given by the magnetic flux as
\begin{equation}
  \frac{k}{2}\sum_{i=1}^Nn_i=\frac{k}{2}\sum_{i=1}^N\tilde{n}_i=
  \frac{kn}{2}>0\ .
\end{equation}
Our strategy is to first obtain classical solutions which would account
for the states with lowest energy (after quantization) for a given positive
$U(1)_b$ charge given above, and then study the small fluctuations
with higher energy. We will see that the latter modes can be treated
perturbatively in $\frac{1}{k}$. From Table 1, the lowest
energy states with positive $U(1)_b$ charge are given by creating
states only with operators $\bar{B}^{\dot{1}},\bar{B}^{\dot{2}}$, or
only with $A_1,A_2$, in their s-waves. The two cases can be analyzed in a
completely same way. They are annihilated by different combinations of
supercharges. Here we consider the first sector only.

Turning on the gauge fields and
$B_{\dot{a}}$, the equations of motion for $A_\mu$, $\tilde{A}_\mu$
are given by
\begin{equation}\label{gauss}
  \frac{k}{4\pi}\epsilon^{\mu\nu\rho}F_{\nu\rho}=-i\sqrt{-g}
  \left(\bar{B}^{\dot{a}}D^\mu B_{\dot{a}}-
  D^\mu\bar{B}^{\dot{a}}B_{\dot{a}}\right)
\end{equation}
where $\epsilon^{t\theta\phi}=1$. $\theta$ and $\phi$ denote the standard
spherical coordinates for the unit 2-sphere.
The equation of motion for $B_{\dot{a}}$ on $S^2\times \mathbb{R}$
is given by
\begin{equation}\label{scalar-eom}
  \hspace*{-0.5cm}\left(\!D^\mu D_\mu\!-\!\frac{1}{4}\right)B_a
  \!=\!\frac{4\pi^2}{k^2}\left[3(B_b\bar{B}^b)^2B_a\!+\!3B_a(\bar{B}^bB_b)^2
  \!-\!2B_b\bar{B}^cB_c\bar{B}^bB_a\!-\!2B_a\bar{B}^bB_c\bar{B}^cB_b
  \!-\!2B_b\bar{B}^bB_a\bar{B}^cB_c
  \right]\ .
\end{equation}
Our solution has nonzero uniform magnetic fields on $S^2$ in the first $U(1)$ among $U(1)^N$ in each $U(N)$, and s-wave of $B_{\dot{a}}$ is nonzero in the $11$ component, where
the first (second) $1$ denotes the first component in the anti bi-fundamental
(bi-fundamental) of first (second) $U(N)$ group. The Gauss' law demands
\begin{equation}
  B_{\dot{a}}=b_{\dot{a}}e^{-it/2}
\end{equation}
where the complex constants $b_{\dot{a}}$ satisfy
\begin{equation}\label{background-gauss}
  |b_1|^2+|b_2|^2=\frac{kn}{4\pi}\ .
\end{equation}
We also set $A_t=0$. One finds that this solution also satisfies the
scalar equation of motion (\ref{scalar-eom}) with the right hand side
from the potential vanishing. The positive
frequency $\omega=\frac{1}{2}$ of our solution implies that $\bar{b}^a$
modes are to be regarded as creation operators after quantization. These
states belong to protected short multiplets, which is easy to see as they
are obtained by $SU(2)$ actions on states of the form
$(b_1^\dag)^{kn}|0\rangle$ \cite{Dolan:2008vc}.

We now investigate fluctuations of all fields
around the above background in the leading order in $\frac{1}{k}$.
The modes run over $A_a,B_{\dot{a}}$, $\psi_a,\chi_{\dot{a}}$ in the
matter fields as well as the vector fields $A_\mu$, $\tilde{A}_\mu$
which turn out to couple to some of the matter fields.

We start by considering bosonic fluctuations. We first consider
the fluctuations $\delta A_a$ in bifundamental of $U(N)\times
U(N)$, which can be considered seperately since they do not mix in the leading
order with other fields in the background with nonzero $B_a$.
The mode $(\delta A_a)_{11}$ or $(\delta A_a)_{ij}$ do not couple to magnetic
field or the background scalar $B_a$ in the leading quadratic order, where
$i,j=2,3,\cdots,N$
are fundamental/anti-fundamental indices for $U(N\!-\!1)\times U(N\!-\!1)$.
These are simply expanded with the spherical harmonics, which
is the same as the weakly coupled theory without monopoles.

The modes $(\delta A_a)_{1i}$ or $(\delta A_a)_{i1}$, which
couple to $\pm n$ units of magnetic charges, are expanded with monopole
spherical harmonics. Monopole spherical harmonics are labeled by the total
angular momentum $j=\frac{|n|}{2},\frac{|n|}{2}\!+\!1,\frac{|n|}{2}\!+\!2,\cdots$
and the Cartan $j_3=m$. From the kinetic and the conformal mass terms,
one obtains
\begin{equation}\label{A-kinetic}
  \frac{d(\delta A_a)}{dt}\frac{d(\delta\bar{A}^a)}{dt}
  -\left[\left(j+\frac{1}{2}\right)^2-\frac{n^2}{4}\right]
  \delta A_a\delta\bar{A}^a\ ,
\end{equation}
where $\delta A_a$ denotes either $1i$ or $i1$ component.
The leading contribution of this fluctuation is also present in the potential.
In the potential, it may appear either by directly fluctuating $A_a,\bar{A}^a$
which are explicit in (\ref{matter-action}), or via fluctuations of the
composite fields $\sigma$, $\tilde\sigma$. The latter possibility yields
no leading terms, quadratic in $\delta A_a$:
$\delta\sigma,\delta\tilde\sigma$ coupling to $A_a$'s obviously starts
from sextic fluctuations with coefficients $\frac{1}{k^2}$ while those
coupling to $B_{\dot{a}}$ always comes with a factor of the
background fields $(\tilde\sigma B_{\dot{a}}-B_{\dot{a}}\sigma)$ or
its conjugate which is zero since $\sigma=\tilde\sigma$ commute with
background $B_a$. Thus we only study the direct fluctuations. One first finds
a factor
\begin{equation}\label{A-potential}
  -\frac{n^2}{4}\delta A_a\delta\bar{A}^a
\end{equation}
from the term coupling to $\sigma,\tilde\sigma$. One should also
consider the potential coming from the superpotential: the term
relevant for the fluctuations $\delta A_a$ is
\begin{equation}\label{A-superpotential}
  -\frac{4\pi^2}{k^2}{\rm tr}\left[(B_{\dot{1}}A_aB_{\dot{2}}-
  B_{\dot{2}}A_aB_{\dot{1}})(\bar{B}^{\dot{2}}\bar{A}^a\bar{B}^{\dot{1}}
  -\bar{B}^{\dot{1}}\bar{A}^a\bar{B}^{\dot{2}})\right]\ .
\end{equation}
The contribution from this term is zero. This is easy to see,
since by an $SU(2)$ internal rotation $b_{\dot{a}}$ can be rotated
to satisfy either $b_1=0$ or $b_2=0$.
Since the superpotential  is $SU(2)$ invariant and acquires nonzero
contribution only when the two scalars are both nonzero, the vanishing of
the fluctuation in this case is obvious.\footnote{In subsection
2.3, we will find a subtle contribution from the superpotential for backgrounds
with more general monopoles.} Combining all, one finds that the last term
in (\ref{A-kinetic}) and (\ref{A-potential}) cancel that the frequencies
of these modes are given by
\begin{equation}\label{A-spectrum}
  \omega^2=\left(j+\frac{1}{2}\right)^2\ .
\end{equation}
In particular, the
spectrum $\omega=\pm\left(j+\frac{1}{2}\right)$ is crucial since some
highest weight states, satisfying $j=j_3$, should saturate the BPS energy
bound
\begin{equation}
  \epsilon\geq R+j_3=j_3+\frac{1}{2}
\end{equation}
where $R$ is the R-charge of $\mathcal{N}=2$ supersymmetry.
This was also found in \cite{Kim:2009wb}.

We then turn to the fluctuations of the scalars $\delta B_a$ in $1i$ and
$i1$ component, where again the first and second indices are for the first
and second of $U(N)\times U(N)$. (Again, `diagonal' modes with $11$ and
$ij$ are trivially expanded with spherical harmonics.)
It is convenient to decompose
\begin{equation}
  \delta B_a=z_a\delta\phi+\epsilon_{ab}\bar{z}^b\delta\varphi\ ,
\end{equation}
where we define $\epsilon_{12}=-\epsilon_{21}=1$ and
$b_a\equiv\sqrt{\frac{kn}{4\pi}}z_a$, satisfying
$|z_1|^2+|z_2|^2=1$. From its $SU(2)$ index structure, the mode $\delta\varphi$
does not directly couple to the background scalar $b_a$ in the leading (quadratic)
order and only couples to it via nonzero $\sigma,\tilde\sigma$. This is the
same as the fluctuations $\delta A_a$ above, leading to the same result
(\ref{A-spectrum}).

Finally in the bosonic sector, we consider
off-diagonal fluctuation $\delta\phi$.
We denote by $\delta\phi^\pm$ the $i1$ and $1i$ components of the first and
second gauge group, respectively.
$\delta\phi^\pm$ couples to $\pm n$ units of fluxes. We also denote by
$\phi_0=\sqrt{\frac{kn}{4\pi}}e^{-it/2}$
the background field.  By taking a glance at the off-diagonal
components of (\ref{gauss}) and (\ref{scalar-eom}),
it turns out that one has to expand the off-diagonal
$A_\mu,\tilde{A}_\mu$ together with $\delta\phi$.
Let us denote by $\delta A_\mu$ and $\delta \tilde{A}_\mu$ the $1i$
component in the adjoint of the first/second gauge group, respectively.
From Gauss law one obtains
\begin{equation}
  \star D\delta A=\frac{2\pi}{k}\left[\!\frac{}{}\!|\phi_0|^2\delta
  A-i\left(\bar\phi_0 D\delta\phi^+\!-\!d\bar\phi_0 \delta\phi^+\right)\right]
  ,\ \star D\delta\tilde{A}=\frac{2\pi}{k}
  \left[\!\frac{}{}\!-|\phi_0|^2\delta\tilde{A}
  -i\left(d\phi_0\delta\bar\phi^-\!-\!\phi_0D\delta\bar\phi^-\right)\right].
\end{equation}
Inserting the value of $|\phi_0|^2$, we rewrite it as
\begin{equation}\label{linear-gauss}
  \left(\star D-\frac{n}{2}\right)\delta A=
  -\frac{2\pi i}{k}\left(\bar\phi_0 D\delta\phi^+-d\bar\phi_0 \delta\phi^+
  \right)\ ,\ \
  \left(\star D+\frac{n}{2}\right)\delta\tilde{A}=
  -\frac{2\pi i}{k}\left(d\phi_0\delta\bar\phi^- -\bar\phi_0D\delta\bar\phi^-
  \right)\ .
\end{equation}
$\delta A$, $\delta\tilde{A}$ are taken
to be of same order as $\frac{1}{\sqrt{k}}\delta\phi^\pm$.
$D$ acts on fluctuations according to their charges:
$D=d-iA^{(0)}$ on $\delta A$, $\delta\tilde{A}$, $\delta\phi^+$ and
$D=d+i A^{(0)}$ on $\delta\phi^-$, where $A^{(0)}$ provides uniform $U(1)$
magnetic field with $n$ units of flux.
Expanding the scalar equation of motion, one finds
\begin{eqnarray}\label{linear-eom}
  &&\left(D^\mu D_\mu-
  \frac{1}{4}\right)\delta\phi^+
  +i\phi_0\left(D^\mu\delta A_\mu\right)
  +2i\partial_\mu\phi_0 \delta A^\mu=0\nonumber\\
  &&\left(D^\mu D_\mu-
  \frac{1}{4}\right)\delta\bar\phi^-
  +i\bar{\phi_0}(D^\mu\delta\tilde{A}_\mu)
  +2i\partial_\mu\bar{\phi_0}\delta\tilde{A}^\mu=0\ .
\end{eqnarray}
Note that the potential does not contribute since it vanishes for a
single complex scalar. All covariant derivatives here and below are
associated with the background magnetic field, and when necessary,
it is also spatially covariantized as well.

To proceed, we act $D\star$ on the first equation of
(\ref{linear-gauss}) and obtain
\begin{equation}
  i\frac{n}{2}{\rm vol}_{S^2}\wedge\delta A-\frac{n}{2}D\star \delta A=
  -\frac{2\pi i}{k}\left(
  \bar\phi_0 D\star D\delta\phi^+-d\star d\bar\phi_0 \delta\phi^+\right)=
  -\frac{2\pi i}{k}\bar\phi_0 \left(D\star D\delta\phi^+
  +\frac{1}{4}{\rm vol}_3 \delta\phi^+\right)
\end{equation}
where we used $\star^2=-1$, $d\star
d\bar\phi_0=-\frac{1}{4}\bar\phi_0{\rm vol}_3$, and
\begin{equation}
  \left(D^2\delta A\right)_{\mu\nu\rho}=3!D_{[\mu}D_\nu\delta A_{\rho]}
  =\frac{3!}{2}[D_{[\mu},D_\nu]\delta A_{\rho]}=-i\frac{3!}{2}F_{[\mu\nu}\delta
  A_{\rho]}=-i\left(F\wedge \delta A\right)_{\mu\nu\rho}\ .
\end{equation}
Since $D\star D\delta\phi=-{\rm vol}_3D^\mu D_\mu\delta\phi$,
$D\star\delta A=-{\rm vol}_3 D^\mu \delta A_\mu$, one obtains
\begin{equation}
  i\frac{n}{2}\delta A_0+\frac{n}{2}D^\mu\delta A_\mu=
  \frac{2\pi i}{k}\bar\phi_0 \left(D^\mu D_\mu\delta\phi^+
  -\frac{1}{4}\delta\phi^+\right)\ .
\end{equation}
Multiplying $\phi_0$ on both sides, one obtains
\begin{equation}\label{reproduce-eom}
  i\frac{n}{2}\phi_0\delta A_0+\frac{n}{2}\phi_0 D^\mu\delta A_\mu=
  i\frac{n}{2}\left(D^\mu D_\mu\delta\phi^+
  -\frac{1}{4}\delta\phi^+\right)\ ,
\end{equation}
which is exactly the scalar equation of motion for $\delta\phi^+$ in
(\ref{linear-eom}). Similar manipulation with the second equation of
(\ref{linear-gauss}) yields the $\delta\bar\phi^-$ equation in
(\ref{linear-eom}). From this finding, we conclude that it suffices
for us to solve (\ref{linear-gauss}) only.

One can easily check the following gauge invariance of
(\ref{linear-gauss}):
\begin{eqnarray}\label{linear-gauge}
  &&\delta A\mapsto \delta A+D\epsilon\ ,\ \ \delta\phi^+\mapsto
  \delta\phi^+-i\phi_0\epsilon\nonumber\\
  &&\delta\tilde{A}\mapsto \delta\tilde{A}+D\tilde{\epsilon}\ ,\ \
  \delta\bar\phi^-\mapsto
  \delta\bar\phi^- -i\bar\phi_0\tilde\epsilon\ .
\end{eqnarray}
This comes from the \textit{linearized} off-diagonal part of the
$U(N)\times U(N)$ gauge transformation, which at this order
does not change the background magnetic field. A convenient gauge
is $\delta\phi^+\!=\!0$ and $\delta\bar\phi^-\!=\!0$, which resembles
the `unitary gauge' in spontaneously broken gauge theories with Higgs
fields. The resulting equation is
\begin{equation}\label{self-dual}
  \left(\star D-\frac{n}{2}\right)\delta A=0\ ,\ \
  \left(\star D+\frac{n}{2}\right)\delta\tilde{A}=0
\end{equation}
where $D$ in both equations is $D=d-iA^{(0)}$. These equations
(with ordinary derivative replacing $D$) are known as odd dimensional
self-duality equations \cite{Townsend:1983xs}, which find their
natural appearances in gauged supergravity theories.

We solve these equations
by expanding with monopole vector spherical harmonics. This is most
easily done by reformulating the problem on $\mathbb{R}^3$, after
\textit{formally} defining Euclidean-like variable $\tau\equiv it$. ($\tau$ is imaginary below.)
One obtains
\begin{equation}
  \star D \Psi =-i\frac{n}{2}\Psi\ ,\ \
  \star D\tilde\Psi =i\frac{n}{2}\tilde\Psi
\end{equation}
where $\Psi,\tilde\Psi$ are $\delta A,\delta\tilde{A}$ on
Euclidean $S^2\times \mathbb{R}$: $(\Psi)_\tau=-i(\delta A)_t$,
$(\Psi)_{\theta,\phi}=(\delta A)_{\theta,\phi}$, etc. Defining $r=e^\tau$
and rescaling fields with
$\frac{1}{r}$, namely $\Psi_r=\frac{1}{r}\Psi_\tau$, etc., one obtains
\begin{equation}
  \vec{\nabla}\times\vec\Psi=-i\frac{n}{2r}\vec\Psi\ ,\ \
  \vec{\nabla}\times\vec{\tilde\Psi}=-i\frac{n}{2r}\vec{\tilde\Psi}
\end{equation}
on $\mathbb{R}^3$.
Actually this is the same expression as that appearing in
the computation of superconformal index in \cite{Kim:2009wb}.
There the 1-loop determinant
$\vec{D}\times\delta\vec{A}-i[\sigma,\delta\vec{A}]-\vec{D}\delta\sigma$
over the bosonic part of vector multiplet is computed. The gauge chosen in
\cite{Kim:2009wb} was the Coulomb gauge, but the differential operator
becomes the same if one chooses $\delta\sigma=0$ gauge instead.

We consider the two equations together below, where
the upper/lower sign denotes the case with $\Psi,\tilde\Psi$, respectively.
We look for configurations on
$S^2\times\mathbb{R}$ with frequency $\omega$ and angular momentum $j$, given by
\begin{equation}
  \vec\Psi,\vec{\tilde\Psi}=
  \frac{1}{r^\omega}\left(a_+\vec{C}^+_{jm}+a_-\vec{C}^-_{jm}
  +a_0\vec{C}^0_{jm}\right)\ .
\end{equation}
$\vec{C}^\lambda_{jm}$ are monopole vector spherical harmonics with $n$ units
of flux on $S^2$. $\lambda=+1,0,-1$ for $j\geq\frac{|n|}{2}\!+\!1$, and
$\lambda=+1,0$ for $j=\frac{|n|}{2}$. Finally, $\lambda=+1$ for
$j=\frac{|n|}{2}\!-\!1$. See Appendix B.2 of \cite{Kim:2009wb}. Note that
$\frac{1}{r^\omega}=e^{-\omega\tau}=e^{-i\omega t}$
is the energy factor. For $j\geq\frac{|n|}{2}\!+\!1$, one finds
\begin{equation}\label{matrix3}
  \left(\begin{array}{ccc}\omega\pm\frac{n}{2}&0&s_+\\
  0&\omega\mp\frac{n}{2}&s_-\\s_+&-s_-&\mp\frac{n}{2}\end{array}\right)
  \left(\begin{array}{c}a_+\\a_-\\a_0\end{array}\right)=0\ ,
\end{equation}
where $s_\pm=\sqrt{\frac{j(j\!+\!1)-q^2\pm q}{2}}$ with $q=\frac{n}{2}$,
and the upper and lower sign is for $\Psi$ and $\tilde\Psi$, respectively.
Nonzero solution exists when the determinant of
$3\times 3$ matrix is zero,
\begin{equation}
  \frac{n}{2}\left[\left(\omega\pm\frac{1}{2}\right)^2-
  \left(j+\frac{1}{2}\right)^2\right]=0\ .
\end{equation}
There are two independent solutions for $\Psi$ with $\omega=-(j+1)$ and
$j$, and for $\tilde\Psi$ with $\omega=-j$ and $j+1$.
For $j=\frac{n}{2}$, we lose the mode $\vec{C}^-$. Also in this case
$s_-=0$ and $s_+=\sqrt{q}$. The equations for $\vec\Psi,\vec{\tilde\Psi}$
are
\begin{equation}\label{det-2by2}
  \left(\begin{array}{cc}\omega\pm\frac{n}{2}&s_+\\
  s_+&\mp\frac{n}{2}\end{array}\right)
  \left(\begin{array}{c}a_+\\a_0\end{array}\right)=0\ .
\end{equation}
The solution exists when $\omega=\mp(\frac{n}{2}+1)=\mp(j+1)$.
Finally, for $j=\frac{n}{2}\!-\!1$ (when $n\geq 2$),
only the mode $\vec{C}^+$ remains. The equation reduces to
\begin{equation}
  \left(\omega\pm\frac{n}{2}\right)a_+=0\ ,
\end{equation}
which has solution if $\omega=\mp\frac{n}{2}=\mp(j+1)$.

Note that, in our gauge which simplified the analysis,
the parallel modes $\delta\phi$ and the orthogonal modes
$\delta\varphi$ apparently look different. We would now like
to rewrite the solutions such that the background-dependence of
$\delta A_\mu,\delta\tilde{A}_\mu,\delta B_a$ fluctuations can be addressed
in a simple manner. To this end, we try to gauge transform the modes $\delta
A_\mu$ and $\delta\tilde{A}_\mu$ back to $\delta\phi$ using (\ref{linear-gauge})
and make the latter look similar to the $\delta\varphi$ modes.
Certainly this is possible for all modes with $j\geq\frac{n}{2}$ since there
are corresponding $\epsilon,\tilde\epsilon$ scalar modes which do this
job.\footnote{As summarized in \cite{Kim:2009wb}, action of $D$ does not
change the values of $j,m$ or the frequency $\omega$ in this case.}
The modes from $\delta A_\mu$ with
$j\geq\frac{n}{2}\!+\!1$, having frequency $\omega=-(j\!+\!1),j$ go to
$\delta\phi^+\!=\!-i\phi_0\epsilon$ with frequency
$\omega=\mp\left(j\!+\!\frac{1}{2}\right)$ due to the multiplication of $\phi_0$.
Similarly, the modes from $\delta\tilde{A}_\mu$ go to $\delta\bar\phi^-$ with
frequencies $\omega=\pm\left(j\!+\!\frac{1}{2}\right)$.
This is identical to the spectrum of $\delta\varphi$ with $j\geq\frac{n}{2}\!+\!1$.
For $j=\frac{n}{2}$, one needs modes with
$\omega=\pm\left(\frac{n}{2}\!+\!1\right)$ in $\delta\phi^\pm$ to match the
spectrum of $\delta\varphi$, which would translate to the modes in
$\delta A_\mu$ with frequency $\omega=\frac{n}{2},-\frac{n}{2}\!-\!1$ and
$\delta\tilde{A}_\mu$ with frequency $\omega=-\frac{n}{2},\frac{n}{2}\!+\!1$.
However, from our analysis in (\ref{det-2by2}), only the latter frequencies for $\delta A_\mu$,
$\delta\tilde{A}_\mu$ exist in the spectrum.
We understand it as considering $\delta B_a$ having universal spectrum (namely,
$\omega\!=\!\pm\left(j\!+\!\frac{1}{2}\right)$ for $j\geq\frac{n}{2}$) and
regard the lacking modes in $\delta\phi$ as constraints
\begin{equation}\label{scalar-constraint}
  [\delta\phi]_{j=\frac{n}{2},~\omega=j\!+\!\frac{1}{2}}\sim
  \bar{b}^a\left[\delta B_a\right]_{j=\frac{n}{2},~\omega=j\!+\!\frac{1}{2}}
  =0
\end{equation}
for all $1i$ and $i1$ components of $\delta B_a$. Since the background
$\bar{b}^a$ carries charges under $U(1)\times U(1)\subset U(N)\times U(N)$,
one should understand this constraint either as the $1i$ component in the adjoint
of first $U(N)$, or the $i1$ component in the adjoint of second $U(N)$. Since
the frequency is positive, the constrained creation operator would
be its conjugate after quantization. Finally,
the modes with $j=\frac{n}{2}\!-\!1$ in $\delta A_\mu,\delta\tilde{A}_\mu$
cannot be gauge-transformed to scalars, since they are genuinely vector-like
modes. They remain as the modes in $1i$ components of each $U(N)$.

To finish the analysis of classical solutions, we consider the
off-diagonal fermions. For $\psi_{a\alpha}$, in its $1i$ and $i1$ components
denoted by $\psi_a^+$ and $\psi_a^-$, the equation of
motion is given by
\begin{equation}
  -i\gamma^\mu D_\mu\psi_a^\pm\pm i\frac{n}{2}\psi_a^\pm=0\ .
\end{equation}
The last term comes from the coupling to $\sigma,\tilde\sigma$
background like \cite{Kim:2009wb}. The spatial part of the differential operator
in this equation is the same as that appearing in Appendix B.1 of
\cite{Kim:2009wb}. There are modes with
frequencies $\omega=\pm\left(j\!+\!\frac{1}{2}\right)$ for $j\geq\frac{n+1}{2}$,
and furthermore modes with $\omega=j\!+\!\frac{1}{2}$ for $j=\frac{n-1}{2}$.
The last modes provide creation operators for $\bar\psi^a$.

To obtain the $\chi_a$ equation, one should also study terms obtained
by integrating out $\lambda_\alpha,\tilde\lambda_\alpha$:
\begin{equation}
  -\frac{4\pi i}{k}{\rm tr}\left[
  \left(\psi_a\bar{A}^a-\bar{B}^a\chi_a\right)
  \left(A_b\bar\psi^b-\bar\chi^bB_b\right)\right]
  +\frac{4\pi i}{k}{\rm tr}\left[
  \left(\bar{A}^a\psi_a-\chi_a\bar{B}^a\right)
  \left(\bar\psi^bA_b-B_b\bar\chi^b\right)\right]\ .
\end{equation}
Denoting by $\chi_a^\pm$ the fluctuations in $1i$ and $i1$
components of first and second $U(N)$ group, respectively, the
equation of motion is given by
\begin{equation}\label{dirac-1}
  -i\gamma^\mu D_\mu\chi_a^+ -i\frac{n}{2}\chi_a^++
  \frac{4\pi i}{k}\chi^+_b\bar{b}^bb_a=0\ ,\ \
  -i\gamma^\mu D_\mu\chi_a^- +i\frac{n}{2}\chi_a^-
  -\frac{4\pi i}{k}b_a\bar{b}^b\chi_b^-=0\ .
\end{equation}
We decompose the fermions as (recall $b_a=\sqrt{\frac{kn}{4\pi}}\ z_a$)
\begin{equation}
  \chi_a^\pm=z_a \xi^\pm + \epsilon_{ab}\bar{z}^b\zeta^\pm\ .
\end{equation}
The Dirac equations for $\xi^\pm$ and $\zeta^\pm$ are
\begin{equation}\label{dirac-2}
  -i\gamma^\mu D_\mu\xi^\pm\pm i\frac{n}{2}\xi^\pm=0\ ,\ \
  -i\gamma^\mu D_\mu\zeta^\pm\mp i\frac{n}{2}\zeta^\pm=0\ .
\end{equation}
The spectrum of $\zeta^\pm$ is same as that of $\psi_a$ above, since the
differential operator in the second equation of (\ref{dirac-2}) is the same
as that for $\chi_a$ in \cite{Kim:2009wb}. However, the sign of the second
term in $\xi^\pm$ equation is flipped from that in \cite{Kim:2009wb}, due
the the last terms in (\ref{dirac-1}). The modes with
$\omega=\pm\left(j\!+\!\frac{1}{2}\right)$ for $j\geq\frac{n\!+\!1}{2}$
remain the same, but the modes for $j=\frac{n-1}{2}$ come with negative
frequency $\omega=-\left(j\!+\!\frac{1}{2}\right)
=-\frac{n}{2}$ due to this sign change. The last modes create
$\xi^\pm$ instead of $\bar\xi^\pm$.

We would again like to express the spectrum of $\xi^\pm,\zeta^\pm$ in
a way such that background dependence is addressed simply. We add by
hand modes in $\xi^\pm$ with $j=\frac{n-1}{2}$ and
frequency $\omega=j\!+\!\frac{1}{2}$, just like $\zeta^\pm$.
Then the whole spectrum of $\chi_a$ is the same as that obtained in
\cite{Kim:2009wb}. The additional mode we inserted is eliminated by the following
fermionic constraint
\begin{equation}
  [\xi^\pm]_{j=\frac{n-1}{2},~\omega=j+\frac{1}{2}}\sim
  \bar{b}^a\left[\chi_{a}^\pm\right]_{j=\frac{n-1}{2},~\omega
  =j+\frac{1}{2}}=0\ .
\end{equation}
The constraint is either in the $1i$ or $i1$ component, which is regarded
as adjoint components in the first and second $U(N)$ gauge group, respectively,
due to the multiplication of $\bar{b}^a$. Finally, there are left-over
modes $\sim\bar{b}^a\chi_a^\pm$ with
$j=\frac{n-1}{2}$ and $\omega=-\left(j\!+\!\frac{1}{2}\right)=-\frac{n}{2}$
in the $1i$ of first $U(N)$ and $i1$ of second $U(N)$, respectively,
from the analysis in the above parenthesis.

Table 2 summarizes adjoint modes and constraints. As for the
indices, $(1i)_1$ denotes the $1i$'th component in the adjoint
representation of the first gauge group (subscript), for instance.
\begin{table}[t]\label{adjoint-modes}
$$
\begin{array}{c|c|ccc|c}
  \hline{\rm creation\ operator}&{\rm nature}
  &\epsilon&h_3&\epsilon-h_3-j_3&({\rm indices})_{\rm group}\\
  \hline(\delta A_\mu)_{j=\frac{n}{2}\!-\!1}\ \
  (\delta\tilde{A}_\mu)^\ast_{j=\frac{n}{2}\!-\!1}&{\rm bosonic\ states}&
  \frac{n}{2}&0&j\!-\!j_3\!+\!1&(1i)_1\ \ (i1)_2\\
  b_a(\delta\bar{B}^a)_{j=\frac{n}{2}}&{\rm bosonic\ constraints}&
  \frac{n}{2}&0&j-j_3&(i1)_1\ \ (1i)_2\\
  \bar{b}^a(\chi_a)_{j=\frac{n\!-\!1}{2}}&{\rm fermionic\ states}&
  \frac{n\!+\!1}{2}&1&j-j_3&(1i)_1\ \ (i1)_2\\
  b_a(\bar\chi^a)_{j=\frac{n\!-\!1}{2}}&{\rm fermionic\ constraints}&
  \frac{n\!-\!1}{2}&-1&j\!-\!j_3\!+\!1&(i1)_1\ \ (1i)_2\\
  \hline
\end{array}
$$
\caption{adjoint states and constraints ($\omega\!<\!0$, creation operators)}
\end{table}

\subsection{Quantization and the superconformal index}

Quantizing the modes obtained in the previous subsection at weak-coupling $\frac{N}{k}\rightarrow 0$, 
including the background modes $b_a$, one can identify the Fock space. From this one can compute
the partition function. Although the last computation is straightforward, in this
paper we only compute the superconformal index.

In the quantum theory (at weak coupling), one can compute the symplectic 2-form on the phase space to obtain normalized oscillators. We do not do this explicitly here, but simply regard the modes with positive/negative
frequencies as annihilation/creation operators, respectively.\footnote{However, we should mention the modes from gauge fields in Table 2, since it is not clear \textit{a priori} if $\omega\gtrless 0$ corresponds to annihilation/creation operators for these modes. The symplectic form is proportional to
\begin{equation}
  -\frac{k}{4\pi}\sqrt{g_{S^2}}\ \epsilon^{\mu\nu}\left(\delta A_\mu\wedge\delta A_\nu-\delta\tilde{A}_\mu\wedge\delta\tilde{A}_\nu
  \right)\ ,\nonumber
\end{equation}
with opposite signs for two gauge fields. From the definition of $\vec{C}^{+1}_{jm}\!\sim\!(\vec{D}+i\hat{r}\times\vec{D})Y_{jm}$ given in \cite{Kim:2009wb}, one can show that the modes from $\delta A_\mu$ with $\omega<0$ are indeed creations, while that from $\delta\tilde{A}_\mu$ with $\omega>0$ are annihilations.} The background variables $b_a$
are regarded as annihilation operators. The Gauss' law
(\ref{background-gauss}) on the background now constrains the
summation of the occupation numbers by
\begin{equation}
  kn=b_a^\dag b_a
\end{equation}
after correct normalization. However, all other small fluctuations
now participate on the right hand side of this equation as well. The
right hand side is the generator of the global part of the
$U(1)\times U(1)\subset U(N)\times U(N)$ gauge transformation. Since
only the difference of two $U(1)$ couples nontrivially to matters,
it may be viewed either as the generator of first $U(1)$ or minus
the generator of second $U(1)$. The presence of $kn$ on the left
hand side implies that the $U(1)$ gauge singlet condition should be
imposed in the presence of the background $U(1)$ charge, which is
$-kn$ or $+kn$ from the viewpoint of the first/second gauge group,
respectively. The gauge invariance condition for $U(N\!-\!1)\times
U(N\!-\!1)$ is applied in the standard way. To impose these
conditions later, it is convenient to introduce $2N$ chemical
potentials $\alpha_i,\tilde\alpha_i$ ($i=1,2,\cdots,N$) for the
$U(1)^N\times U(1)^N\subset U(N)\times U(N)$ color charges. Our
convention is that states with positive charges are weighted by positive
powers of $e^{-i\alpha_i}$.

The partition function of the free theory is easily calculated by
first considering the single particle partition function, namely
that over the modes. We define it by
\begin{equation}
  z(x,x^\prime,y_1,y_2,\zeta,\alpha,\tilde\alpha)=
  {\rm tr}\left[x^{\epsilon+j_3}
  (x^\prime)^{\epsilon-h_3-j_3}y_1^{h_1}y_2^{h_2}\zeta^{2j_3}
  e^{-i\sum_{i=1}^N(\alpha_iq_i+\tilde\alpha_i\tilde{q}_i)}\right]
\end{equation}
where $q_i,\tilde{q}_i$ are $U(1)^N\times U(1)^N$ charges.
The one particle index is obtained by setting $\zeta\!=\!-1$:
\begin{equation}
  f(x,y_1,y_2,\alpha,\tilde\alpha)=z(x,x^\prime,y_1,y_2,-1,\alpha,\tilde\alpha)=
  {\rm tr}\left[(-1)^Fx^{\epsilon+j_3}y_1^{h_1}y_2^{h_2}
  e^{-i\sum_{i=1}^N(\alpha_iq_i+\tilde\alpha_i\tilde{q}_i)}\right]\ .
\end{equation}
The dependence on chemical potential $x^\prime$ disappears in this
limit. These quantities are also called partition functions or
index over `letters.'

Let us compute the contribution to $f$ from various modes. Firstly, the
modes in $11$ or $ij$ (for $i,j\neq 1$) of
$U(N)\times U(N)$ are basically the same as those of conventional free
field theory, considered in \cite{Bhattacharya:2008bja}. The one
particle index from bi-fundamental modes is given by
\begin{equation}\label{bif-no-flux}
  e^{-i(\alpha_1-\tilde\alpha_1)}f^+(x,y_1,y_2)+\sum_{i,j=2}^N
  e^{-i(\alpha_i-\tilde\alpha_j)}f^+(x,y_1,y_2)
\end{equation}
where
\begin{equation}
  f^+=\left(\sqrt{\frac{y_1}{y_2}}+\sqrt{\frac{y_2}{y_1}}\right)
  \frac{x^{\frac{1}{2}}}{1-x^2}-
  \left(\sqrt{y_1y_2}+\sqrt{\frac{1}{y_1y_2}}\right)
  \frac{x^{\frac{3}{2}}}{1-x^2}\ ,
\end{equation}
and the anti bi-fundamental index is given by
\begin{equation}\label{anti-no-flux}
  e^{i(\alpha_1-\tilde\alpha_1)}f^-(x,y_1,y_2)+\sum_{i,j=2}^N
  e^{i(\alpha_i-\tilde\alpha_j)}f^-(x,y_1,y_2)
\end{equation}
where
\begin{equation}
  f^-=\left(\sqrt{y_1y_2}+\sqrt{\frac{1}{y_1y_2}}\right)
  \frac{x^{\frac{1}{2}}}{1-x^2}-
  \left(\sqrt{\frac{y_1}{y_2}}+\sqrt{\frac{y_2}{y_1}}\right)
  \frac{x^{\frac{3}{2}}}{1-x^2}\ ,
\end{equation}
as explained in \cite{Bhattacharya:2008bja}.

To evaluate the contribution to the single particle index from $1i$, $i1$ of $U(N)\times
U(N)$ (where $i\!\neq\!1$), we follow the decomposition of charged modes in the previous
subsection, into bi-fundamentals and adjoints. The bifundamental/anti-bifundamental
part of the index is simply given by that of \cite{Kim:2009wb},
since all the modes are same in two cases. This is given by
\begin{equation}\label{1i-flux}
  \sum_{i=2}^N\left[e^{-i(\alpha_1-\tilde\alpha_i)}x^{n}f^+(x,y_1,y_2)+
  e^{i(\alpha_1-\tilde\alpha_i)}x^nf^-(x,y_1,y_2)\right]
\end{equation}
plus
\begin{equation}\label{i1-flux}
  \sum_{i=2}^N\left[e^{-i(\alpha_i-\tilde\alpha_1)}x^{n}f^+(x,y_1,y_2)+
  e^{i(\alpha_i-\tilde\alpha_1)}x^nf^-(x,y_1,y_2)\right]\ .
\end{equation}
Finally, we sum over the finite number of adjoint
modes which are either in $1i$ or $i1$ components of one of the two $U(N)$
gauge groups. The necessary information is summarized in Table 2. One finds
the following indices (bosonic/fermionic constraints appear with $\mp$ signs,
respectively)
\begin{eqnarray}
  {\rm gauge\ fields}\!&\!:\!&\!
  \sum_{i=2}^N xx^\prime\left[(x^\prime)^{n\!-\!2}+
  (x^\prime)^{n\!-\!3}x+\cdots+x^{n\!-\!2}\right](e^{-i(\alpha_1-\alpha_i)}
  +e^{-i(\tilde\alpha_i-\tilde\alpha_1)})\nonumber\\
  {\rm bosonic\ constraint}\!&\!:\!&\!
  -\sum_{i=2}^N\left[(x^\prime)^{n}+(x^\prime)^{n\!-\!1}x+\cdots+x^{n}\right]
  (e^{-i(\alpha_i-\alpha_1)}+e^{-i(\tilde\alpha_1-\tilde\alpha_i)})\nonumber\\
  {\rm fermionic\ states}\!&\!:\!&\!
  -\sum_{i=2}^N x\left[(x^\prime)^{n\!-\!1}+
  (x^\prime)^{n\!-\!2}x+\cdots+x^{n\!-\!1}\right](e^{-i(\alpha_1-\alpha_i)}
  +e^{-i(\tilde\alpha_i-\tilde\alpha_1)})\nonumber\\
  {\rm fermionic\ constraint}\!&\!:\!&\!
  \sum_{i=2}^N x^\prime\left[(x^\prime)^{n\!-\!1}+
  (x^\prime)^{n\!-\!2}x+\cdots+x^{n\!-\!1}\right](e^{-i(\alpha_i-\alpha_1)}
  +e^{-i(\tilde\alpha_1-\tilde\alpha_i)}).
\end{eqnarray}
Adding the first/third lines, and also the second/fourth lines, one
observes a vast cancelation. The final answer for the adjoint single
particle index is
\begin{equation}\label{adjoint-flux}
  f^{\rm adj}(x,\alpha,\tilde\alpha)=-x^n\sum_{i=2}^N\left[
  e^{-i(\alpha_1-\alpha_i)}+e^{-i(\tilde\alpha_i-\tilde\alpha_1)}+e^{-i
  (\alpha_i-\alpha_1)}+e^{-i(\tilde\alpha_1-\tilde\alpha_i)}\right]\ ,
\end{equation}
which is in perfect agreement with the result of \cite{Kim:2009wb}.

Let us define the sum of (\ref{bif-no-flux}), (\ref{anti-no-flux}),
(\ref{1i-flux}), (\ref{i1-flux}) to be $f^{\rm
matter}(x,y_1,y_2,\alpha,\tilde\alpha)$. The full index is obtained
from one particle index by the multi-particle (or
Plethystic) exponential of $f^{\rm matter}+f^{\rm adj}$. At this
point we impose the $U(1)$ and $U(N-1)\times U(N-1)$ singlet
conditions with background $U(1)$ charge. The final result is
\begin{eqnarray}
  I(x,y_1,y_2)&=&\int\frac{d\alpha_1d\tilde\alpha_1}{(2\pi)^2}
  e^{ikn(\alpha_1\!-\!\tilde\alpha_1)}
  \frac{1}{[(N\!-\!1)!]^2}\prod_{i=2}^N\left[\frac{d\alpha_id\tilde\alpha_i}
  {(2\pi)^2}\right]\prod_{i<j}\left[2\sin\frac{\alpha_i\!-\!\alpha_j}{2}
  \right]^2\left[2\sin\frac{\tilde\alpha_i\!-\!\tilde\alpha_j}{2}\right]^2
  \nonumber\\
  &&\hspace{1cm}\times\exp\left[\sum_{p=1}^\infty\frac{1}{p}\left(
  f^{\rm matter}(x^p,y_1^p,y_2^p,p\alpha,p\tilde\alpha)+
  f^{\rm adj}(x^p,p\alpha,p\tilde\alpha)\right)\right]\ .
\end{eqnarray}
This was used in \cite{Kim:2009wb} to reproduce the large $N$ supergravity
index.

\subsection{Generalizations and open problems}

One can try to generalize the analysis in the previous two subsections.
Firstly, one can take more magnetic charges to be nonzero
$n_i,\tilde{n}_i\neq 0$ for some $i\geq 2$, while still satisfying $H=\tilde{H}$.
We discuss some aspects of the semi-classical analysis and point out some
puzzles, or subtleties. Secondly, one can consider the case $H\neq\tilde{H}$, which is claimed
to be present during comparison with supergravity \cite{Kim:2009wb}.
We have very little to say about this case here, apart from modest comments
at the end of this subsection.

We first consider the case in which $n_i\!=\!\tilde{n}_i$ for $i=1,2,\cdots,N$
where more than one pairs of fluxes are nonzero. For simplicity, we take all
of them to be non-negative. Among $H=\{n_i\}$, some of
them can be identical. If the fluxes are given by
\begin{equation}\label{general-flux}
  \stackrel{\underbrace{p_1,p_1,\cdots,p_1}\ >}{N_1\ \ }\
  \stackrel{\underbrace{p_2,\cdots,p_2}\ >}{N_2\ \ }\cdots
  \stackrel{>\ \underbrace{p_f,\cdots,p_f}}{\ \ \ N_f}
\end{equation}
where $N_1\!+\!N_2\!+\!\cdots\!+\!N_f\!=\!N$, the gauge symmetry
$U(N)\times U(N)$ is broken to $\prod_{i=1}^fU(N_i)\times U(N_i)$.
Again one can obtain classical solutions which will account for the lowest
energy states. We take all $U(1)^N\times U(1)^N$ magnetic
fluxes in (\ref{general-flux}) to be uniform on $S^2$, set $A_t=0$, and
restrict the scalars $B_{\dot{a}}$ to be s-waves and also to be block-diagonal
in $\prod_{i=1}^f U(N_i)\times U(N_i)$. We take
\begin{equation}
  B_{\dot{a}}=b_{\dot{a}}e^{-it/2}\ ,
\end{equation}
where $b_{\dot{a}}$ are constant block-diagonal matrices.
To solve the Gauss' law, they are subject to the following condition
\begin{equation}\label{background-gauss-2}
  \frac{kp_i}{4\pi}{\bf 1}_{N_i}=\bar{b}^ab_a\ ,\ \
  \frac{kp_i}{4\pi}{\bf 1}_{N_i}=b_a\bar{b}^a
\end{equation}
in each of the $U(N_i)\times U(N_i)$ block.
Inserting the above relation to the scalar equation of motion
(\ref{scalar-eom}), one finds that it is satisfied as well.

The constraint (\ref{background-gauss-2}) from the Gauss' law may be
solved as follows. In each block-diagonal sector, the matrices $b_a$
can be diagonalized with unitary matrices $U_a,V_a$ as
\begin{equation}
  b_a=U_a^\dag\mathcal{D}_aV_a\ .
\end{equation}
The constraints can then be written in the $i$'th block as
\begin{equation}
  U^\dag_2\mathcal{D}_2\mathcal{D}_2^\dag U_2=
  U^\dag_1(\frac{kp_i}{4\pi}-\mathcal{D}_1\mathcal{D}_1^\dag)U_1\ ,\ \
  V^\dag_2\mathcal{D}_2^\dag\mathcal{D}_2V_2=
  V^\dag_1(\frac{kp_i}{4\pi}-\mathcal{D}_1^\dag\mathcal{D}_1)V_1\ .
\end{equation}
One finds that $U_2=S_{N_i}U_1$, $V_2=S_{N_i}V_1$, where $S_{N_i}$
denotes an element of permutation of eigenvalues.
Thus $b_1$ and $b_2$ can be simultaneously diagonalized by
$U(N_i)\times U(N_i)$. Denoting the eigenvalues of
$\mathcal{D}_a$ by $\lambda_{an}$ ($n=1,2,\cdots,N_i$), one obtains
\begin{equation}\label{diagonalized}
  |\lambda_{1n}|^2+|\lambda_{2n}|^2=\frac{kp_i}{4\pi}\ .
\end{equation}
There still remains $S_{N_i}$ permutation symmetry acting on $N_i$
doublets $(\lambda_{1n},\lambda_{2n})$ of eigenvalues.

Before discussing the excitations from this solution, we first try to quantize
the lowest energy states. One can do this in two approaches. One can either
start from the diagonalized variables (\ref{diagonalized}) with
permutation symmetry, or work directly with block-diagonal variables satisfying
(\ref{background-gauss}). The first way is essentially the quantization of
moduli space discussed in, say, \cite{Aharony:2008ug}.

We start by considering the first method. Each pair of eigenvalues
$(\lambda_1,\lambda_2)$, correctly normalized as creation/annihilation
operators after quantization, are constrained to satisfy
$\lambda_1^\dag\lambda_1+\lambda_2^\dag\lambda_2=p_ik$,
constraining the sum of occupation numbers to be $p_i k$. The
single particle partition function is given by
($y$ is conjugate to $SU(2)$ Cartan)
\begin{equation}
  z(x,y)=x^{\frac{p_i k}{2}}
  \left(y^{p_i k}+y^{p_i k\!-\!2}+\cdots+y^{-p_i k}\right)\ .
\end{equation}
The full $N_i$-particle partition function $Z_{N_i}(x,y)$ in $i$'th block
is given by
\begin{eqnarray}\label{lowest-graviton}
  Z(x,y,\nu)&\equiv&\sum_{N=0}^\infty\nu^NZ_N(x,y)=
  \exp\left(\sum_{n=1}^\infty\frac{\nu^n}{n}\zeta(x^n,y^n)\right)\\
  &=&\frac{1}{(1-\nu x^{\frac{p_ik}{2}}y^{p_ik})
  (1-\nu x^{\frac{p_ik}{2}}y^{p_ik\!-\!2})\cdots
  (1-\nu x^{\frac{p_ik}{2}}y^{-p_ik})}\ .\nonumber
\end{eqnarray}
Actually this is the dual graviton partition function at lowest energy.
The full partition function is given by the product of partition functions
for all blocks.

We can also quantize this sector by working with $2N^2$ matrix elements
of oscillators $(b_a)^\dag$. For
the $mn$'th matrix element of $(b_a)^\dag$, the partition function
for the single oscillator is
\begin{equation}
  \frac{1}{1-x^{\frac{1}{2}}y^{\pm 1}e^{-i(\alpha_m\!-\!\tilde\alpha_n)}}
\end{equation}
where $\pm 1$ is for $a=1,2$, respectively.
$\alpha_n,\tilde\alpha_n$ are $2N$ chemical potentials conjugate to
the $U(N_i)\times U(N_i)$ color charges. The quantum version of the
constraints (\ref{background-gauss-2}) are given by
\begin{equation}
  p_i k{\bf 1}_{N_i}=:b_ab_a^{\ \dag}:\ ,\ \
  p_i k{\bf 1}_{N_i}=:b_a^{\ \dag}b_a:
\end{equation}
where $:\ :$ denotes normal ordering. The right hand sides of the
two equations are minus the generator of the second $U(N)$ gauge
transformation, and the generator of the first $U(N)$ gauge
transformation, respectively. The presence of left hand sides imply
that gauge invariance has to be imposed with the background electric
charges ($-p_ik{\bf 1}_{N_i}$,$p_ik{\bf 1}_{N_i}$) for
$U(N_i)\times U(N_i)$.
Collecting all, the partition function counting gauge invariant
states is given by
\begin{equation}\label{lowest-gauge}
  Z_{N_i}(x,y)=\int[dU][d\tilde{U}]\
  e^{ikp_i\sum_{n=1}^{N_i}(\alpha_n\!-\!\tilde\alpha_n)}
  \prod_{m,n=1}^{N_i}\frac{1}{(1-x^{\frac{1}{2}}y
  e^{-i(\alpha_m\!-\!\tilde\alpha_n)})(1-x^{\frac{1}{2}}y^{-1}
  e^{-i(\alpha_m\!-\!\tilde\alpha_n)})}
\end{equation}
where the phase factor is due to the background charge, and
the unitary matrices $U,\tilde{U}$ can be diagonalized with
eigenvalues $\{e^{i\alpha_n}\},\{e^{i\tilde\alpha_n}\}$. The
Haar measure appearing in the integral is
\begin{equation}
  [dU][d\tilde{U}]=\frac{1}{(N_i!)^2}
  \left[\frac{d\alpha_n d\tilde\alpha_n}{(2\pi^2)}\right]
  \prod_{m<n}\left(2\sin\frac{\alpha_m\!-\!\alpha_n}
  {2}\right)^2\left(2\sin\frac{\tilde\alpha_m\!-\!\tilde\alpha_n}{2}
  \right)^2\ .
\end{equation}
This is the same as the field theory index in \cite{Kim:2009wb}, which is
actually the partition function since there are no fermions. Again, the full
partition function is given by the product for all blocks.

We turn to discuss the fluctuations of modes around this
background in the leading order in $\frac{1}{k}$.
Contrary to the analysis in previous sections, we encounter a subtlety
starting from the simplest fluctuations $\delta A_a$. We will simply point
this out in this case, leaving a detailed study for the future.

We take the diagonalized background scalars satisfying
(\ref{diagonalized}). Let us consider the fluctuations $\delta A_a$
in bifundamental of $U(N_1)\times U(N_2)$.\footnote{The case with
$\delta A_a$ within a single block can also be obtained if one replaces
$p_2,N_2$ by $p_1,N_1$ below.} After an analysis similar to section 2.1,
the kinetic and mass terms become
\begin{equation}
  \frac{d\delta {A}_a}{dt}\frac{d\delta\bar{A}^a}{dt}
  -\left[\left(j+\frac{1}{2}\right)^2-\frac{(p_1\!-\!p_2)^2}{4}\right]
  \delta A_a\delta\bar{A}^a
\end{equation}
where $j\geq\frac{|p_1\!-\!p_2|}{2}$. Its appearance in the potential
can also be analyzed similarly. From the coupling of $A_a$ to
$\sigma,\tilde\sigma$ in (\ref{matter-action}), one obtains
\begin{equation}
  -\frac{(p_1\!-\!p_2)^2}{4}\delta A_a\delta\bar{A}^a
\end{equation}
in the leading order. One should also consider the potential coming from
the superpotential in this case: from (\ref{A-superpotential}), one obtains (superscripts in $\lambda^1_{am}$, $\lambda^2_{bn}$ refer to first/second blocks)
\begin{equation}
  -\left|\epsilon^{ab}\lambda^1_{am}\lambda^2_{bn}\right|^2
  \delta A_{amn}(\delta A_{amn})^\ast
\end{equation}
where $m=1,2,\cdots,N_1$ and $n=1,2,\cdots,N_2$. This is nonzero unless
$\lambda^1_{am}$ and $\lambda^2_{an}$ are proportional as $SU(2)$ doublets.
Collecting all, one obtains the classical modes of
$(\delta A_a)_{mn}$ with the following frequency:
\begin{equation}
  \omega^2=\left(j+\frac{1}{2}\right)^2+
  |\epsilon^{ab}\lambda^1_{am}\lambda^2_{bn}|^2\ .
\end{equation}
Apparently, the supersymmetric modes, which should satisfy
$\omega\pm\left(j\!+\!\frac{1}{2}\right)$, seem to be allowed only for
specific backgrounds. It is not yet clear to us how to deal with these modes
and, in particular, address the results in \cite{Kim:2009wb} from our
approach.

The analysis in \cite{Kim:2009wb} also demands the existence of monopole
operators with $H\neq\tilde{H}$, for the index to agree with the index from
supergravity. This case is much harder to study in our approach than examples
above, since we even do not know an exact classical solution in this background.
Note that our semi-classical consideration of the exact backgrounds in the
previous subsections resembles the quantization
of moduli space, which is generally useful in dealing with the modes in s-waves.
From the analysis in \cite{Kim:2009wb}, states including monopole operators
with $H\neq\tilde{H}$ carry nonzero spatial angular momenta, $j_3\neq 0$.
Perhaps it may not be effective to consider this case with our approach.

\section{Concluding remarks}

In this paper we studied the spectrum of local operators which involve magnetic
monopole operators. We considered the semi-classical quantization of all
excitations around the exact classical solution, where the latter accounts
for protected chiral operators with lowest energy in the monopole background.
We used our result to reproduce the superconformal index of \cite{Kim:2009wb}
in the simplest monopole background, namely $H=\tilde{H}=(n,0,0,\cdots,0)$.

A motivation of this study was to demystify some results obtained in
\cite{Kim:2009wb}. Since the calculation there involved deforming
the theory (in a way that the index does not change), it was hard to
see what is actually going on physically, despite all the
quantitative agreement with supergravity reported there. A novel
feature was the appearance of degrees of freedom in the adjoint
representation of $U(N)\times U(N)$ in the presence of nonzero flux.
The analysis of this paper shows that this has to do with the
interaction between some matters and gauge fields. It technically
comes from `exceptional' low-lying spherical harmonics with nonzero
monopoles, with total angular momentum $j=\frac{|n|\!-\!1}{2}$ for
spinors and $j=\frac{|n|}{2},\frac{|n|}{2}\!-\!1$ for vectors.

Another finding is that the actual spectrum is subtler than that in the
`deformed' theory, although the difference is guaranteed not to affect the
index. Knowing the actual Hilbert space in the weakly interacting regime,
one can try to systematically develop the
relevant perturbation theory to higher orders in $\frac{1}{k}$. This will
in principle enable us to study the open strings connecting heavy $D0$ branes
(with mass $\sim k$) from the gauge theory. For instance, it may be
interesting to see if one can obtain a useful open spin chain description
for macroscopic open strings ending on $D0$ branes. See also
\cite{Berenstein:2009sa} for a study of open strings ending on membrane giant
gravitons.

\vskip 0.5cm

\hspace*{-0.8cm} {\bf\large Acknowledgements}

\vskip 0.2cm

\hspace*{-0.75cm} We would like to thank Shiraz Minwalla for many
helpful discussions and suggestions, and Amihay Hanany, Ki-Myeong Lee, Sungjay Lee, Soo-Jong Rey,
Riccardo Ricci for discussions. K.M. would like to acknowledge the
generous support of the people of India for research in the basic
sciences.

\end{document}